\documentclass[twocolumn,english,aps,pre,a4paper,floatfix,superscriptaddress]{revtex4}
\usepackage[T1]{fontenc}
\usepackage{color}
\usepackage[latin1]{inputenc}
\usepackage{graphicx}
\usepackage{bm}
\usepackage{epsfig}
\usepackage{rotating}
\usepackage{float}

\usepackage[normalem]{ulem}

\usepackage{dcolumn}
\usepackage{bm}

\begin{document}

\title{Domain walls in vertically vibrated monolayers of cylinders confined in annuli}

\author{Ariel D\'{\i}az-De Armas}
\affiliation{
Grupo Interdisciplinar de Sistemas Complejos (GISC), 
Departamento
de Matem\'aticas, Escuela Polit\'ecnica Superior, Universidad Carlos III de Madrid,
Avenida de la Universidad 30, E-28911, Legan\'es, Madrid, Spain}

\author{Mart\'{\i}n Maza-Cuello}
\affiliation{Soft Matter Sciences and Engineering (SIMM), ESPCI Paris, PSL University, Sorbonne Universite, CNRS, F-75005 Paris, France}

\author{Yuri Mart\'{\i}nez-Rat\'on}
\email{yuri@math.uc3m.es}
\affiliation{
Grupo Interdisciplinar de Sistemas Complejos (GISC), Departamento
de Matem\'aticas, Escuela Polit\'ecnica Superior, Universidad Carlos III de Madrid,
Avenida de la Universidad 30, E-28911, Legan\'es, Madrid, Spain}

\author{Enrique Velasco}
\email{enrique.velasco@uam.es}
\affiliation{Departamento de F\'{\i}sica Te\'orica de la Materia Condensada,
Instituto de F\'{\i}sica de la Materia Condensada (IFIMAC) and Instituto de Ciencia de 
Materiales Nicol\'as Cabrera,
Universidad Aut\'onoma de Madrid,
E-28049, Madrid, Spain}  

\date{\today}

\begin{abstract}
Liquid-crystalline ordering in vertically vibrated granular monolayers confined in annuli of different sizes is examined. The annuli consist of circular cavities with a central circular obstruction. In the absence of the central obstruction cylinders of low aspect-ratio exhibit tetratic order, except for the existence of four defects which restore the symmetry broken by the circular confinement. This behaviour is demanded by topology in systems with strong anchoring properties at the surface.  By contrast, topology dictates that the annular geometry is compatible with a distorted tetratic phase without point defects. However, the effect of restricted geometry and limited size on phases possessing finite anchoring energy at the wall and elastic stiffness leads to different configurations, showing finite ordered regions separated by domain walls.  We argue that highly packed nonequilibrium vibrated granular monolayers respond to geometrical frustration and extreme confinement as corresponding equilibrium systems of particles do, and that the former can be analysed in terms of surface free energies, elastic distortions and defects, much as equilibrium liquid crystals. Therefore, selective confinement of vertically-vibrated monolayers of rods could be used with advantage as a new tool to study the creation and dynamics of various types of defects in ordered systems.
\end{abstract}


\maketitle

\section{Introduction}

The observation that quasi-two-dimensional monolayers of granular spherical 
particles can be 
excited by periodic motion, leading to pattern formation, has been
exploited in the last decades 
\cite{Taguchi,Williams,Eu,Grossman,Noije,Clement,Warr,Warr1,Kudrolli,Olafsen}.
A milestone in the field of ordering in granular media
was the observation that some structural properties of a vibrated monolayer of 
spheres (in particular, crystallization), were similar to those of the ordinary hard-sphere model, a basic model to understand the structure of matter at the nano- and mesoscopic scales in thermal equilibrium \cite{Mulero}. These phenomena 
can be largely understood by applying the standard rules
of equilibrium statistical mechanics \cite{Ojha,Olafsen}.
Many other nonequilibrium phenomena
are observed in periodically agitated granular matter, e.g. the presence of solitons.

Patterns in vibrated monolayer of particles can be found by 
suitably choosing the control parameters of the
external drive imposed on the system. In this way, vertically-vibrated
granular rods have recently been shown
to exhibit interesting spatial patterns that resemble liquid-crystalline
phases in two-dimensional systems of anisotropic particles subject to thermal
fluctuations \cite{Narayan,Galanis1,Galanis2,Aranson1,Aranson2,Mueller,us,us1}.
Therefore, the inherently non-equilibrium granular arrangements 
of anisotropic grains, when continually excited externally \cite{Melo,Miller}, 
reach steady-states with particle arrangements
which show typical characteristics of liquid crystals: local and global
nematic ordering \cite{Narayan,Galanis1,Mueller}, competing elastic bulk and 
surface energies \cite{Galanis2}, together with typically non-equilibrium
effects \cite{Aranson1,Aranson2,us}. It is tempting to describe the behaviour in these
systems in terms of entropic arguments based on volume exclusion, since forces 
between particles are absent except when they collide, and volume-filling 
concepts become relevant.
Recently the phase diagram of equilibrium
hard rectangles has been explored by Monte Carlo simulation and seen
to exhibit many similarities, even quantitative, with the steady-state
arrangements of granular rods confined in circular cavities \cite{Mueller}.
We must remind ourselves that the behaviour of hard particles driven 
by thermal fluctuations are exclusively controlled by entropy and particle
overlap statistics. This connection between thermal and nonthermal (granular) matter is intriguing.

More recently Gonz\'alez-Pinto et al. \cite{us1,GP}
have analysed defect formation in granular experiments
of metallic rods when their aspect ratio is small. In this regime 
\textit{tetratic} arrangements, where particles form a fluid monolayer but with
their long axes pointing along one of two perpendicular, equivalent, 
directors, are easily stabilised. As dictated by topological considerations,
extended tetratic phases, when confined to a circular cavity, restore
the four-fold symmetry of the director by creating four point defects with
a total topological charge of +4. This is indeed what is observed in the
granular monolayers, showing that these non-equilibrium systems respond
to geometrical frustration in the same way as their thermal-equilibrium
counterparts. Even more, by assuming that these four defects behave as
point defects of charge $+1$ and that they interact with each other 
following elastic theory (i.e. through logarithmic potentials mediated by
the elastic stiffness coefficient of the tetratic phase), the spatial
fluctuations of these defects can be measured and used to extract an
elastic constant, which turns out to be of the same order of magnitude as
typical two-dimensional elastic constants of liquid crystals \cite{GP}.

The picture therefore emerges that, at least in some range of external
parameters, vertically-vibrated monolayers of granular rods 
share many properties with corresponding systems governed
by thermal fluctuations. The cause for this apparent connection
is presently unknown: only some partial ideas and basic relations can be
advanced. We are still in the process of collecting as much evidence as
possible on these similarities. 

In this article we present yet another 
example where granular matter seems to respond to external constraints
in the same way as thermal matter. The system is an annulus, i.e.
a circular cavity with a central circular obstacle
confining the granular particles to a geometry essentially different to that
of a simple
circular cavity. In this case, as discussed below, topology does not
require the formation of point defects. This is indeed observed in our
experiments, but only in part: due to the severe
size restrictions, quantified in terms of the ratio of particle length
and cavity radius, another type of defects, namely (non-point-like) {\it domain walls}
(i.e. extended regions) that separate regions with different liquid-crystalline ordering, are formed. 
These extended regions appear due to the severe size restrictions, quantified in terms of the ratio 
of particle length and cavity radius, and are similar to domain walls
in materials where domains with different orientation of a 
non-scalar order parameter coexist (see e.g. Ref. \cite{Heras} where domain walls separating
uniaxial nematic phases with different orientations were observed in circular cavities
using Monte Carlo simulation).

These structures (which are not required by topology as will be explained below) are created because
elastic, wall effects and intrinsic particle
ordering all compete to somehow satisfy a `maximum entropy principle'.
This is at least the paradigm imposed by equilibrium statistical mechanics,
which again seems to operate, at a presently unknown level, in 
vertically-vibrated monolayers of granular rods (the very definition of bulk, wall and elastic
free energies is not clear as these concepts belong to the realm of equilibrium physics).
On top of that, this connection is a potentially interesting result, as it would mean
that granular matter can be used to easily create and control particle
orientations and defects of various kinds by setting up
particular geometric constraints by way of external boundaries, at 
accessible --macroscopic--
time and length scales. By contrast, this process
is much more difficult to implement using soft matter based, for example, 
on --nanoscopic-- liquid-crystal materials. 

Shortly after our work on granular rods confined in annuli was started,
G\^arlea et al. \cite{Mulder} presented results on a similar system,
but using virus particles (i.e. of colloidal size), which are
subject to Brownian (thermal) motion. Computer simulations were also presented
to corroborate the experimental results. Remarkably, the
structures found by G\^arlea et al. are qualitatively similar to those
obtained by us, although the range of parameters is different. 
Again this demonstrates the hitherto unknown relation
between two systems, granular and thermal, that are governed by 
very different physics at radically different length and time scales.
We will discuss the similarities between our results and those obtained 
by G\^arlea et al. in Sec. \ref{Conclusions}. 

The article is organised as follows. In the following section
we expand on the theoretical background
needed to frame the discussion into the correct context. 
The experiment is then discussed in some detail,
together with the analytical tools used. A section on results
follows where all observations, histograms and distribution functions are
presented. The article ends with a section on discussion and conclusions.

\section{Motivation and theoretical background}

Hard rectangles of sufficiently low aspect ratio $\kappa=L/D$, where
$L$ is the length and $D$ is the width, exhibit an exotic nematic phase
called the tetratic phase.  The orientational distribution function
$h(\varphi)$ of the tetratic phase has four-fold symmetry: it presents 
equal-height peaks at angles $\varphi=0$, $\pi/2$, $\pi$ and $3\pi/2$,
where the origin of angles ($x$-axis in the lab frame) 
is taken to point along one of the directors. This means that, in an
extended tetratic phase, two equivalent directors can be found at a
relative angle of $90^{\circ}$ with respect to each other. Particles
diffuse, as it corresponds to a nematic fluid, but they orient 
along two possible perpendicular directions with higher probability.

The equilibrium phase diagram of hard rectangles has been analysed using
SPT by Schlacken et al. \cite{Schlacken} and more in depth by 
Mart\'{\i}nez-Rat\'on et al. \cite{us2}. 
The stable phases in the phase
diagram are all spatially uniform, since non-uniform phases were not explored.
They are: isotropic (I), uniaxial nematic (N$_u$) and tetratic nematic 
(N$_t$) phases. Tetratic phase stability
occurs for aspect ratios $\kappa$ less than $2.62$. For
larger aspect ratios the stable nematic phase is the uniaxial nematic phase.
The tetratic phase is stable at rather large values of packing fractions,
values which would otherwise correspond to a non-uniform phase, such
as columnar, smectic or crystal. This point may lead one to think that 
the tetratic phase should, in a more complete treatment including
spatially nonuniform order, be preempted by one of these phases. However,
further analysis has shown that this is not the case. 
Inclusion of higher-order orientational correlations via an
effective third-order virial coefficient \cite{us} leads to an 
isotropic-tetratic transition line which is shifted to lower packing fractions.
A consequence of this is that the range of aspect ratios where the tetratic phase
is stable is considerably enlarged up to $3.23$, meaning that three-body 
correlations are crucial in the tetratic phase. There are reasons to 
believe that inclusion of even higher-order correlations will further
expand the stability of the tetratic phase. In fact, recent Monte Carlo 
simulations \cite{Mueller} have shown that the tetratic phase is stable 
for $\kappa\alt 7$,
at values of packing fractions which are too low for the crystal and
other nonuniform phases to be thermodynamically stable.

M\"uller et al. \cite{Mueller} have investigated the relation between the 
equilibrium system of hard rectangles, analysed by means of Monte Carlo simulation, 
and monolayers of granular particles made of cylinders, which project on the horizontal plane approximately
as rectangles. Their experimental setup is similar to ours (to be explained in more detail later):
cylinders were confined into a thin circular cavity which leaves a 
small vertical space so that, when vertically agitated, cylinders 
move in the horizontal directions via collisions, without overlapping.
Uniaxial nematic and tetratic nematic
order parameters were defined in order to identify different types of order. 
Phases were identified as
a function of projected aspect ratio and packing fraction. Overall qualitative
agreement was found between both phase diagrams. In some cases, for example 
in the value of
limiting aspect ratio for tetratic stability, the agreement was even
quantitative. This intriguing similarity still awaits for an explanation.

Gonz\'alez-Pinto et al. \cite{us1} have extended the analysis of this system, confirming
some aspects of the work by M\"uller et al. but casting some doubts in the
case of long cylinders. Even though finite-size effects may be at work in this
case, long cylinders do not seem to develop uniaxial ordering but rather a
kind of patched nematic with no global ordering. Nonetheless, the results for
short cylinders were confirmed. Also, the tendency for clustering of groups
of particles was examined and contrasted with results from Monte Carlo
simulations. Clustering is much stronger in the dissipative system \cite{us1}, 
which is readily understood as particles tend to freeze when they collide, thus
generating an extra ordering tendency. Leaving aside this and other aspects,
the two systems, dissipative and equilibrium, look remarkably similar
in their ordering properties.
\\
\begin{figure*}
\begin{center}
\includegraphics[width=0.90\linewidth,angle=0]{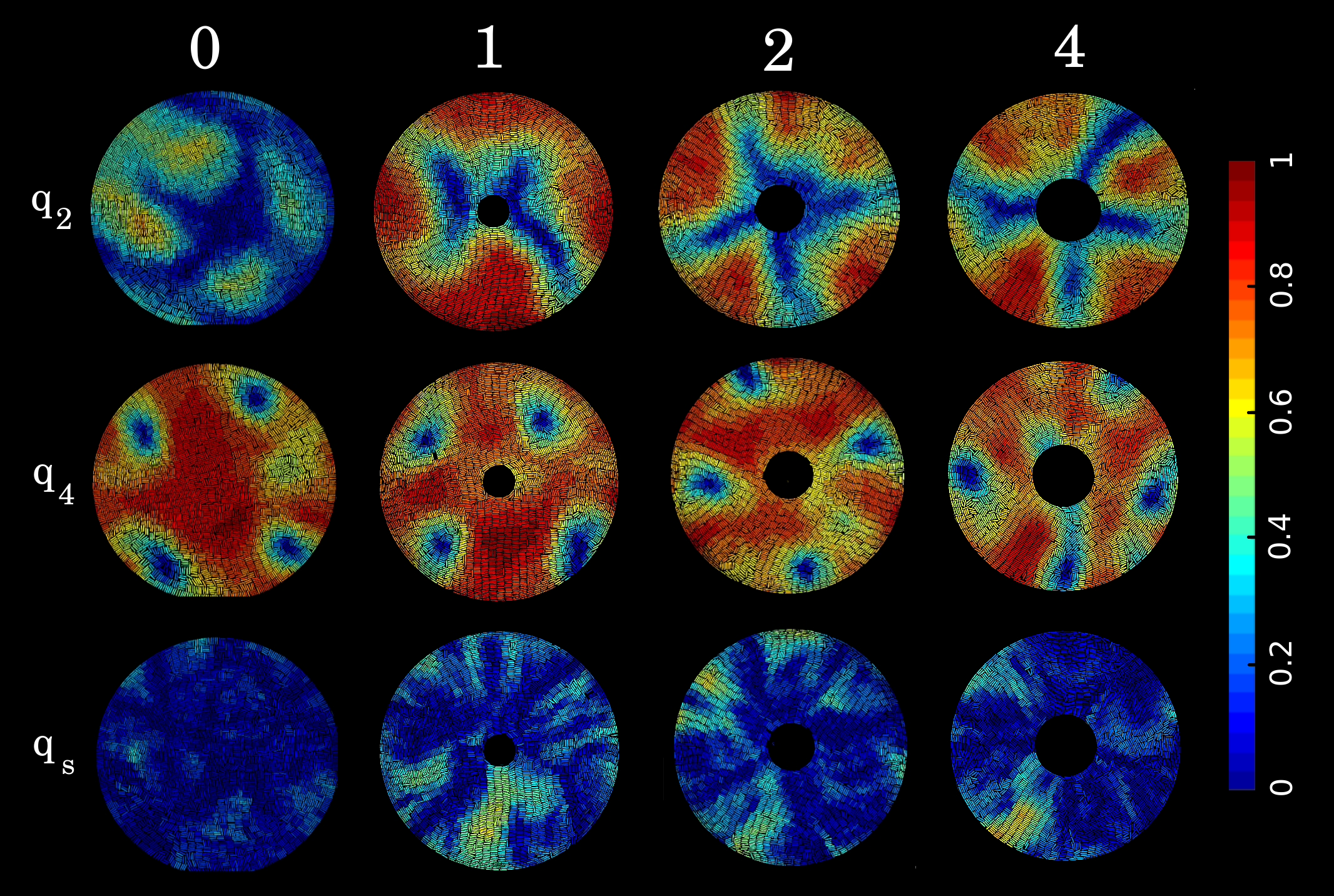}
\caption{\label{fig2} Values of the $q_2$ (uniaxial), $q_4$ (tetratic) and
$q_s$ (smectic) order parameters on each particle, and 
for the four situations investigated in this work. 0: no obstacle;
1: small obstacle; 2: intermediate obstacle; and 4: large obstacle. 
In each case the configuration has been chosen more or less at random
during the course of the respective experiments.
Colour code is shown in the vertical bars. In all experiments
particles had $\kappa=4$ and packing fraction was set to $\eta\simeq 0.75$.
	Values for the other parameters were taken as mentioned in Sec. \ref{analysis}.} 
\end{center}
\end{figure*}

In additional work, Gonz\'alez-Pinto et al. \cite{GP} have made the observation that,
when the tetratic arrangement of the granular system
prevails, the imposed circular geometry of the cavity forces the monolayer 
to create four defects, located close to the inner boundary of the 
container and arranged on average at the corners of a square. See Fig. \ref{fig2},
panel 0--$q_4$ (here we use a matrix style to address each image, with 
column $=\{0,1,2,4\}$ and row $=\{q_2,q_4,q_s\}$. The meaning of these
labels will be explained in Section \ref{Results}). The figure 
shows an instantaneous configuration of the particles, colour-coded 
according to the value of the tetratic
order parameter $q_4$ (to be defined later). The packing fraction is $\eta\simeq 0.75$. 

The presence of the four defects is revealed by the low value of
the order parameter $q_4$, which is high in the region occupied by the tetratic
phase. Topological arguments can be used to explain this configuration.
The notion is based on the so-called {\it geometrical frustration}:
due to the boundary conditions imposed on a system, which otherwise
develops some specific local order in bulk, the order cannot propagate 
to the entire system. In our case, the tetratic order, which imposes 
a four-fold orientational order locally (C$_4$ symmetry in the nomenclature of group theory), 
cannot satisfy the circular
geometry of the cavity, and the continuity of the tetratic field must be
broken in some regions. The excitation of defects is frequent in 
liquid-crystalline materials whose local order-parameter field can be
elastically distorted (with an associated price in elastic free energy),
but the presence of surfaces with some specific geometry may be
incompatible with a distorted order-parameter field because of the high
surface free energy that should be paid to avoid the creation of defects,
even at very low temperature (low energy).

Therefore, we are facing a problem that can be quantified using
the Euler theorem of topology applied to the experimentally-obtained tetratic field \cite{Bowick}. 
In the present context the theorem 
states that the total topological charge of the system must satisfy the
equation $\sum_i{\cal Q}_i=p\chi$, where $p$ is the symmetry order of the
phase (in the case of the tetratic, $p=4$), while $\chi$ is the 
Euler characteristic of the volume. 
In the case of a circular cavity consisting of a disc, $\chi = 1$. $Q_i$ is the `topological charge' of 
each individual defect present in the volume, a number that reflects the symmetry of the order parameter 
around the defect. The Euler theorem gives the total charge of the system, not the number or topological charge 
of the defects present. These properties depend on other considerations based on the competition between free 
energy contributions of different origin: bulk, wall, elastic and defect core energies. In any case the system 
will minimise the free energy by choosing the minimum number of defects that can restore the broken symmetry 
induced by the confining boundary.
In a circular cavity the presence of four defects
restores the symmetry broken by the circular geometry of the cavity since the 
total charge $+4$ satisfies the constraint imposed by the Euler theorem for a medium with C$_4$ symmetry. 

It is remarkable that the steady-state structures reached by a driven system of 
dissipative particles follows general laws based on the order-parameter concept,
meant to describe behaviour in systems in thermal equilibrium. With a view to further 
exploring the previous findings for the circular cavity, 
in this article we describe the results found in
experiments where the granular rods are confined in a cavity with a totally different
topology, namely a circular annulus: a circular cavity with a central hole. 
In this case the Euler characteristic is $\chi=0$,
meaning that the total topological charge inside the volume should be zero. In fact, as
discussed later, the integrity of the director field can be maintained, without the presence of
any singularities, by distorting the field around the central hole without ever creating any
conflict with the wall contour. We will see that in this geometry the topological requirements
are not entirely fulfilled by the granular system,
due to reasons that will become clear later.

\section{Experimental setup}

In the experiment 
cylinders made of nonmagnetic steel with
length 4 mm and width 1 mm (aspect ratio $\kappa=L/D=4$) are placed inside
a cylindrical cavity of radius $R=7$ cm ($R/L=17.5$). The two planar,
horizontal plates
of the cavity are close enough so that cylinders have a free height of
1.8 mm. Therefore cylinders cannot pass each other and constitute an
effective monolayer. The cavity is mechanically agitated at frequency 
$\nu=90$ Hz. The effective acceleration of the system can be measured in
terms of the dimensionless parameter $\Gamma=a_0\nu^2/g$, where
$a_0$ is the amplitude of the vibration, and $g$ is the acceleration of
gravity. In our experiments $\Gamma\simeq 2$--$3$. The upper lid of the
cavity is made of transparent plastic so that a zenithal DSLR
camera aligned along the cavity axis 
can record the time evolution of the particles.
The cavity was carefully aligned with the horizontal by using a highly
sensitive bubble level. Vibration of the system
was induced by connecting the container to an electromagnetic
shaker, which allows for the frequency $\nu$ and effective amplitude 
$\Gamma$ to be controlled. This is extremely
important as the behaviour of the monolayer depends on these two parameters.
Their values were chosen in order to avoid undesired nonequilibrium effects 
such as collective motion or creation of holes.

Particle identification (of position and orientation)
is done using a {\sc Matlab}$^{\textregistered}$ code implemented by the authors. The code locates
the centre of each particle and the angle between the long axis of the
particle and a reference $x$ axis. Distortion due to the curvature effects
is negligible since the camera is at a long distance from the
cavity and works at relatively long focal distance. Uniform illumination of
the cavity is achieved by placing light diffusers surrounding the experiment.
The identification software is very successful as typically only between 1--3\% of
the particles are not correctly identified.

In each experiment the protocol used was the following. First the initial configuration
is prepared by hand: the upper lid of the cavity is removed, and cylinders are placed
on the lower surface, avoiding any overlaps. Then the cavity is covered by the lid and the
packing fraction estimated by taking a picture of the static system and using the identification software.
In case the target density is not achieved particles are removed or added 
according to the difference between actual and target packing fractions. 
Images are taken typically every 15 s, and experiments are run for several hours. In all cases 
the system gets readily ordered after the experiment is started, 
but a completely stable (in the steady-state sense) regime is attained only after an hour or more.

\section{Analysis}
\label{analysis}

Previous work on a similar experimental setup \cite{us} 
obtained different types of particle arrangements, depending mainly
on the packing density of the system: I, where particles are disordered in both
orientations and positions; N$_t$, where particles show fluid behaviour but are oriented on average
along two equivalent, perpendicular directions; and smectic (S), with particles forming fluid layers.

Each of these configurations can be identified by means of a number of order parameters.
In this case, we define three order parameters. On the one hand, two orientational order parameters:
\begin{eqnarray}
q_n=\left<\cos{n\theta}\right>\hspace{0.4cm}\hbox{with}\hspace{0.4cm}n=2,4,
\end{eqnarray}
which probe two- and four-fold symmetries, respectively uniaxial and tetratic 
symmetries.
$\theta$ is the angle between the long axis of a particle and the local alignment
direction $\hat{\bm n}$ (local director). On the other hand, we define the smectic order parameter as
\begin{eqnarray}
q_s=\left<e^{i{\bm q}\cdot{\bm r}}\right>
\end{eqnarray}
(the meaning of the brackets is defined below).
Here ${\bm r}$ is the position of a particle, and ${\bm q}$ a wavevector compatible with the 
cylinder length (in fact a little larger to allow for smectic layer fluctuations). Both these
vectors are referred to the frame defined by the local director.
One essential characteristic of our order-parameter-based description is that $q_2, q_4, q_s$ are
not defined as local fields at a point ${\bm r}$, but on each particle. This is similar to the
Eulerian versus Lagrangian views of the flow field of a fluid. The advantage of this approach is
that visualisation of order in separate configurations is much easier. On the other hand, our
`Lagrangian' approach also allows to obtain average values over the whole cavity in the steady state.

For the definition of all three order parameters the local director 
$\hat{\bm n}$ on each particle is required.
This is obtained by first defining a circular region $C$ of radius $\xi=4L$ centred at each particle, 
and then calculating the $2\times 2$ tensor
\begin{eqnarray}
	Q=\left<2\hat{\bm e}_k\hat{\bm e}_k^T-I\right>=
\frac{1}{M}\sum_{k=1}^M\left(2\hat{\bm e}_k\hat{\bm e}_k^T-I\right),
\end{eqnarray}
which defines the average $\left<\cdots\right>$.
Here $\hat{\bm e}_k$ is the unit column vector along the long axis of 
particle $k$, the sum in $k$ extends over all $M$ 
particles whose centres of mass 
are contained in the region $C$, and $I$ is the $2\times 2$ identity matrix. The eigenvector
of $Q$ associated with the largest eigenvalue defines the local director 
$\hat{\bm n}$ on the particle.

Now local configurations can be identified as follows: 
\begin{itemize}
\item I: $q_2\sim q_4\simeq 0$ and $q_s\simeq 0$
\item N$_u$: $q_2\simeq q_4$ and $q_s\simeq 0$
\item N$_t$: $q_2\ll q_4$ and $q_s\simeq 0$
\item S: $q_2\agt q_4$ and $q_s>0$
\end{itemize}
An essential property of the steady states of this system is that, except at low packing fractions
(a regime which is not explored in detail in the present work), the values of the order
parameters are not the same, not even of the same order, in different regions of the cavity.
Note that, in our experiments for $\kappa=4$, extended uniaxial nematic configurations are not
formed for any value of aspect ratio. 

The local order parameters represent a powerful tool to identify the ordering in the cavity
in space and also in time. In practice we analyse all images taken by the acquisition system,
identifying particles and calculating the three order parameters on each particle. Correspondingly,
three sequences of images are produced, which are colour-coded according to the value of the
order parameter. Examples are shown in Fig. \ref{fig2}, with each
row representing a different order--parameter field. 
The processed images are piled up to produce videos, which are very helpful
to visualise the ordering dynamics, particle motion and evolution of defected structures in the system.
In addition to these order parameters, we have also defined a number of distribution functions 
for the defects inside the cavity, which will be explained in the following section.

\section{Results}
\label{Results}

We have investigated four different systems: no obstacle, `0', small obstacle, `1' (1 cm in diameter), intermediate obstacle, `2' (2 cm in diameter), 
and large obstacle, `4' (4 cm in diameter). 
These labels are used in Fig. \ref{fig2}.
Several experiments with different packing fraction in each case were considered. However, we
concentrate on comparing cases with approximately the same packing fractions and different
obstacle size. The reason is that the interesting density window is relatively narrow: 
(i) for $\eta\alt 0.70$ the tetratic phase is not well developed in the cavity, and therefore 
defects are not well-defined entities along the whole 
time span of the experiments; and (ii) systems with
$\eta>0.75$ are very difficult to prepare manually due to the high density. Therefore, we focus on
systems with $\eta\simeq 0.75$, for which a well structured tetratic configuration 
is developed. 

First, a comment on the wall orientation of the particles is in order. Density-functional theory 
applied on the equilibrium hard-rectangle model (with the restricted-orientation approximation)
has shown \cite{Yuri} that the preferred orientation of rectangles close to a flat wall is planar, 
i.e.  with the long particle axes aligned preferentially parallel to the wall.
This is the most sensible result on theoretical grounds based on packing considerations.
However, as the wall gets curved, planar packing may be less efficient and, for a radius of
curvature larger but of the same order as the particle length, perpendicular alignment
may be more optimal. This change of behaviour has not been proved to occur yet, either
theoretically or by simulation. Given the parameter values used in our experiments, 
we seem to be far from this situation.
Notwithstanding this, one cannot discard perpendicular alignment in the
vibrated granular rods, since strong nonequilibrium phenomena such as non-uniform granular
temperature distributions (with granular temperature defined in terms of average horizontal
kinetic energy)
may be more important than simple packing effects, leading to different alignments. 
Although in most of the experiments to be presented below 
planar alignment obtains, perpendicular orientation has indeed been observed in a few
cases; rationalization of this result and the investigation of its origin 
would require more systematic studies (not performed in this work).
 
\subsection{No obstacle: circular cavity}

This system was studied previously \cite{Mueller,us1,GP}, and is 
here reanalysed as a reference case (note that the experiment was slightly redesigned 
with respect to that of Refs. \cite{us1,GP}; in particular, the cavity is slightly bigger).
In this section we also define some other quantities that will be used in the
other cases.
Fig. \ref{fig2}, column 0, shows a typical configuration in the steady state. All three order parameters
are shown. The high value of the $q_4$ order parameter in a large fraction
of the area, together with the low value of $q_2$ and $q_s$, point to the
formation of a large domain of tetratic symmetry. In addition, there are four regions 
where all three order parameters are depleted, corresponding to regions
where particles are orientationally disordered. These regions are `point'
defects which restore the global symmetry of the tetratic phase. As discussed
previously, this symmetry is broken by the circular cavity, whose symmetry is incompatible 
with the four--fold symmetry of the tetratic phase. The defects are located
at the corners of a square, and are `point--like' in the sense that 
their associated depleted region is confined to a region (the `core'
of the defect) of finite area. This tetratic configuration with
four defects still forces some distortion of the local tetratic directors 
close to the walls between contiguous defects, but the distortion is 
assimilated by the system through the accummulation of `elastic free
energy', without producing additional defective regions.

The position of the four defects fluctuates in time, in a manner that will
be described below. However, they always stay close to the cavity wall. 
In the frame of equilibrium elastic theory, this would be explained by the existence 
of an effective repulsive interaction between the defects. They interact as if they were points particles,
the interaction being mediated by the intervening,
tetratic phase. The interactions can be modelled in terms of logarithmic repulsions, which are explained by elastic theory. The fact that these interactions
are reasonably pairwise and isotropic (i.e. they only depend on distance between pairs
of defects) explains the square configuration for the defects.
Interactions are transmitted through the tetratic medium because
of the stiffness coefficient $K$ of the medium: despite being fluid, the tetratic material
is elastic with respect to local distortion of their (locally perpendicular)
directors. $K$ is the only factor that contributes to the strength of the repulsive interaction. Since it  
is expected to increase with density, the medium should
become increasingly stiff as the packing fraction of the tetratic medium
gets larger. Indeed this is clearly seen in the experiments by simple visual inspection
(not shown here). The dynamics of the
defects can in fact be assimilated to that of four point-like interacting Brownian particles
(which fluctuate due to the presence of the faster-moving granular particles).
These Brownian particles interact through the repulsive logarithmic 
force which depends on the
stiffness coefficient $K$. Details can be found in Ref. \cite{GP}.

In our experiments, defects were identified and located as follows. First, from the $q_4$ map,
which associates a value of $q_4$ to each granular particle, we select 
those particles with an order parameter $q_4<0.4$. This criterion is 
sufficient to isolate separate clusters in the system, which will be 
potential candidates for the point-like defects. In effect, on most occasions 
only four regions are found (otherwise the configuration is discarded), as can be seen in Fig. \ref{fig2}.
The centre of mass of particles that belong to a given isolated cluster is taken
as the defect location.

Typically, in the course of the experiments, the system is prone to 
developing a global rotation, which closely corresponds to a rigid rotation.
However, this rotation does not follow a temporal pattern and resembles
a chaotic motion in time, both in amplitude and in the direction of rotation. 
In order to eliminate this rotation and extract the inherent fluctuations
of the defects, we define an instantaneous reference frame where the 
average azimuthal position of the defects is constant. Motion of the 
defects with respect to this frame should then contain the inherent fluctuations of the
defects. In the remainder of this article it should be understood that
all results are defined in this average reference frame.

In Ref. \cite{GP} it proved useful to characterise the motion of the defects
in terms of distribution functions. We also adopt this approach here and define two functions:
\begin{itemize}
\item $f_1(r)$. This is the {\it radial function}, which gives
the spatial distribution of the defects with respect to their distance $r$
from the centre of the cavity. 
\item $f_2(s)$. This is the {\it inter-defect function}, which gives
the distribution with respect to the relative distance between two 
defects $s$.
\end{itemize}
Both functions are averaged over time. $f_1(r)$ is further averaged 
over the four defects, while $f_2(s)$ is averaged over distinct pairs of
defects. $r$ and $s$ are conveniently scaled by the cavity radius $R$. Since
defects are mostly located close to the cavity wall, one expects that
$f_1(r)$ presents a maximum at a radial distance $r/R\alt 1$, while the
square-like arrangement of the four defects gives rise to 
two peaks in $f_2(s)$, associated with the distance between first
neighbours (along the sides of an imaginary square) and second neighbours
(along the square diagonal), with the two distances having a ratio $\sqrt{2}$.
Analysis of these two functions allows for a proper evaluation of the 
excursions of the defects inside the cavity and the fluctuations of their
spatial arrangement which, in general and as mentioned before, we expect 
it to have a square geometry on average. Since $f_1(r)$ is more informative 
than $f_2(s)$ in the case of the annuli, we only report results on the radial
function.

\begin{figure}
\begin{center}
\includegraphics[width=0.9\linewidth,angle=0]{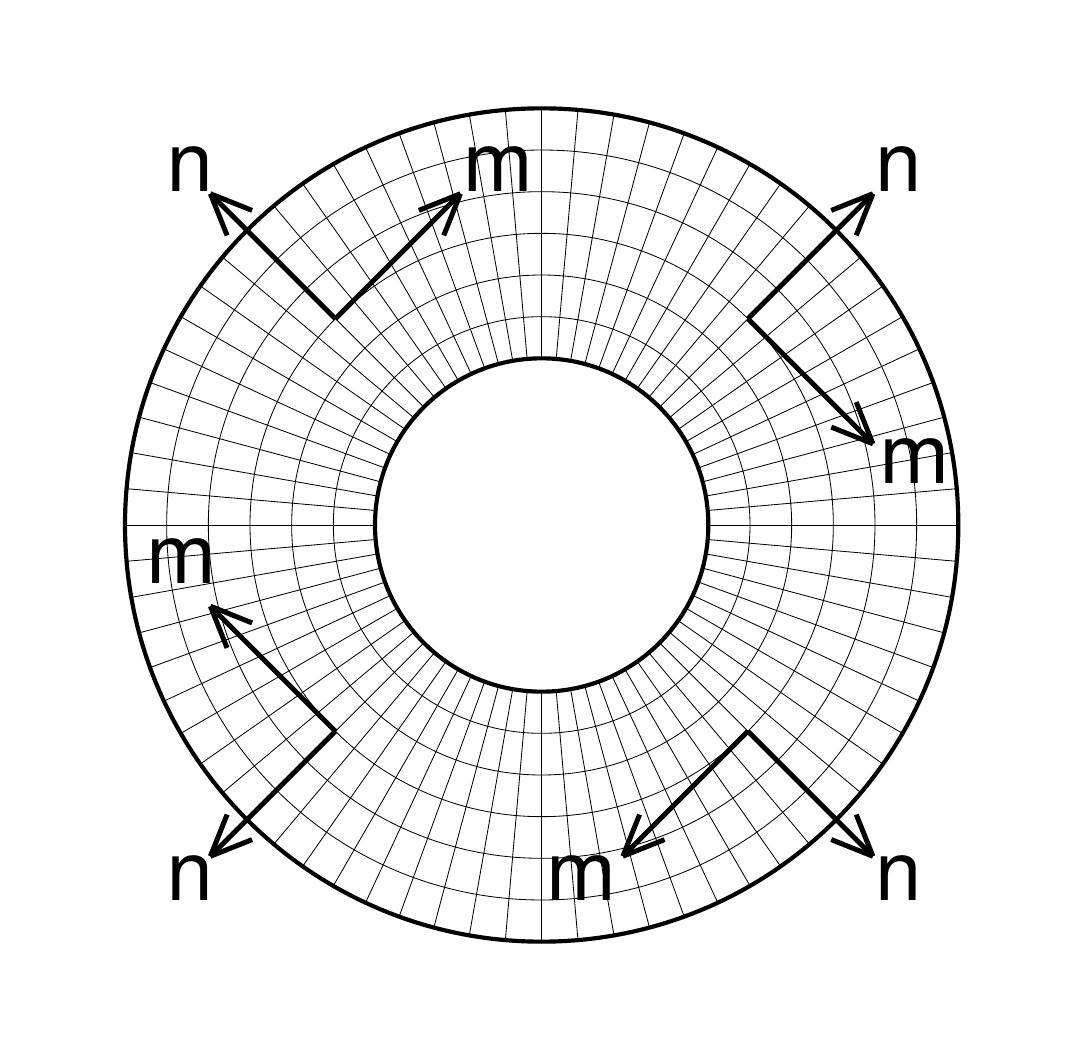}
\caption{\label{distort} Schematic of the director field of
a tetratic configuration inside an annulus. The local orientation of
the directors $\hat{\bm n}$ and $\hat{\bm m}$ is indicated at four
particular locations. The system is invariant under the local symmetry
operations $\hat{\bm n}\to -\hat{\bm n}$ and 
$\hat{\bm m}\to -\hat{\bm m}$. Continuous lines represent director
field lines.}
\end{center}
\end{figure}

\subsection{Annuli}

Having discussed the reference system, where no central obstacle is introduced
in the cavity, we now turn to the annular cavities.  
The topology of the annulus is different from that
of the disc, and in fact is fully compatible with that of the tetratic
phase. Therefore, we should not expect the presence of defects in the cavity
but simply a tangentially distorted tetratic phase, with the two locally perpendicular 
directors $\hat{\bm n}$ and $\hat{\bm m}$ 
showing splay ($\hat{\bm n}$, 
the one pointing in the radial direction) and bend ($\hat{\bm m}$, 
the one along the azimuthal direction) deformations, respectively.
Note that the system is invariant under the symmetry
operations $\hat{\bm n}\to -\hat{\bm n}$ and
$\hat{\bm m}\to -\hat{\bm m}$.
See Fig. \ref{distort} for a schematic representation of a distorted
tetratic directors inside an annulus; distortion avoids the excitation of
defects because the condition of parallel orientation at the boundary 
is perfectly fulfilled without compromising the global symmetry of the system.

Topological arguments were successful in explaining the global properties
of the cavity in the case where no obstacle is present.
However, even though topological arguments preclude the existence of defects 
in the annular cavity, things are more complicated. A central problem is
the finite size of the system. Topology assumes the existence of a continuous
field (the director field), except at point-like regions where the field
is not defined, and states the conditions under which singularities
should occur and the number and characteristics of these singularities.
Another assumption is the {\it strong boundary condition} at the walls, 
i.e. the fact that the director field has to be oriented as dictated by
the wall, with no relaxing or free-energy penalty condition.
This assumption may not be valid in real situations.

To understand this point more clearly, let us introduce the typical lengths of
our system: $L$, the particle length; $R$, the radius
of the cavity; and $R_{\rm obs}$, the radius of the central obstacle.
The continuous field assumption rests on the conditions $R\gg L$ and 
$R_{\rm obs}\gg L$. While the first condition is probably fulfilled in our experiments,
which have $R/L=17.5$ (and the success of topological arguments in the no-obstacle
case strongly supports this), in the case of the small obstacle we have $R_{\rm obs}=0.5$ cm
and consequently $R_{\rm obs}/L=1.25$.
The perimeter of the central
obstacle is covered by only $\sim 2\pi R_{\rm obs}/L\sim 8$ particles 
forming an octagon.
Certainly, the continuum approximation breaks down, and topological
arguments may not be valid in this case.
For the other obstacles we have $R_{\rm obs}/L=2.5$ and $5$, still too small for the continuum approximation to be valid.

The results for the granular system confined in annuli, in comparison
with the reference case with no obstacle,
can be inferred very clearly from the order-parameter maps, Fig. \ref{fig2}.
In this figure typical particle configurations with
the value of the order parameters superimposed are shown.
The column labelled 1 corresponds to the
small obstacle. At first sight, from the $q_4$ map, we may think that the 
four defects, already analysed in the obstacle-free cavity, 
are still present. Indeed, from the $q_4$ maps for the other cases with 
increasing obstacle size, we can draw the same conclusion: spatially 
limited regions with a depleted tetratic order parameter are clearly
excited. The arguments taken from topology are apparently not valid
which, at the end of the day, is not a far-reaching conclusion
as the conditions for the continuum approximation are not fulfilled 
by any of the experiments using an annular cavity.

There are, however, important differences between the circular and annular
cavities. If we forget about topology which, as discussed above, may not be valid
to explain the results, we may regard the obstacle 
as a perturbation to the tetratic field weakening
the long-range interaction between defects, especially
second-neighbour defects (those
located along the diagonal of the square defined by the four defects).
As a consequence, we might expect to see larger
fluctuations in the defect positions: 
excursions of the defects away from
the wall and into the bulk of the cavity will be longer, while the
integrity of the defects as point-like particles will remain intact.
However, this effect is not very clear
when comparing the $q_4$ maps as we move from experiment 1 to 
experiment 4.

\begin{figure}
\begin{center}
\includegraphics[width=0.90\linewidth,angle=0]{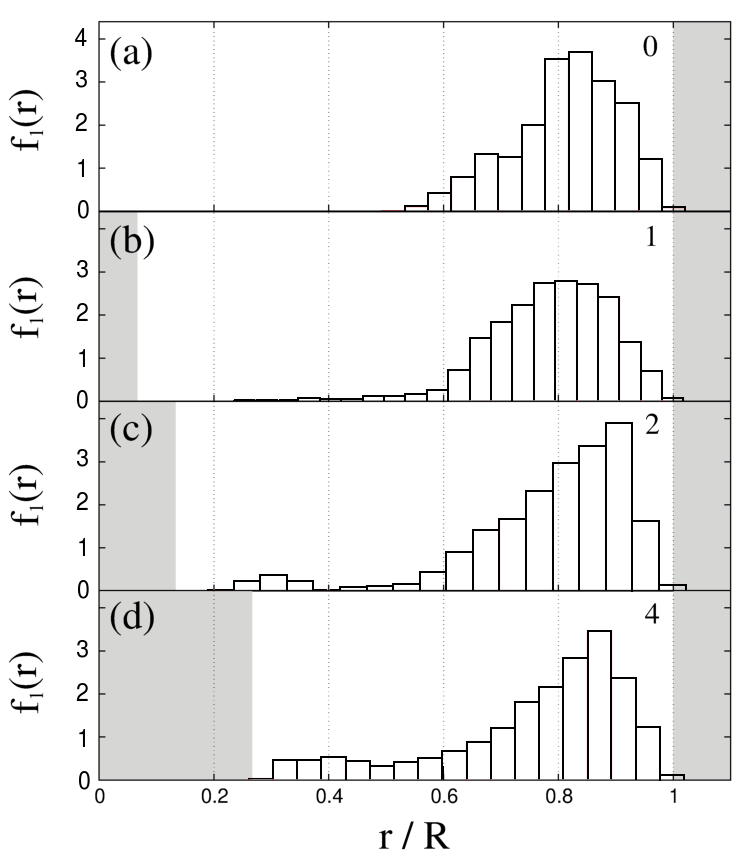}
\caption{\label{fig4} Radial distribution of defect positions for 
(a) no obstacle, (b) obstacles of diameter $d=1$, (c) $d=2$ and
(d) $d=4$. Shaded areas indicate regions that cannot be explored by the
defects (central obstacle at left and region outside cavity at right).}
\end{center}
\end{figure}

The key idea 
comes from a combined interpretation of all three order-parameter
fields. The $q_2$ maps show a clear-cut qualitative
difference between the obstacle-free
case and the rest. In the first case, the order parameter is low, except
for localised, very dynamic regions, which correspond to the
excitation of local uniaxial order, associated with smectic order.
These local structures are short-lived and decay quite rapidly.
However, in the annular experiments, the $q_2$ maps exhibit 
very apparent and persistent structures with a high value of $q_2$. 
On the one hand, there are large regions 
where the $q_2$ order parameter is depleted. The remarkable 
thing is that both regions (with high and low values of $q_2$)
are \textit{spatially complementary}:
The order parameters are indicating the presence of a
cross-shaped structure consisting of `bridges', oriented at relative
angles of $90^{\circ}$, which connect the wall of the container with
the wall of the central obstacle. 
One of these bridges can be seen in column 4 of Fig. \ref{fig2}.
In between these bridges, regions
with a high value of the $q_2$ order parameter are created. By 
looking at the $q_4$ and $q_s$ maps in each case, we can reach the
conclusion that these four regions correspond to stable smectic 
domains. These structures are very robust and persistent in time, 
the more so as the size of the central obstacle becomes larger.

The comparison between the different order-parameter maps not
only provides a hint about the structure of the regions in between
the cross-shaped bridges, but also the structure inside the bridges.
This structure is complex. Close to the cavity walls both $q_2$ and $q_4$ 
are depleted, which indicates that particles are orientationally
disordered (isotropic region); this is reminiscent of the no-obstacle 
situation. But as we move away from this wall towards the central 
obstacle and along the radial direction, we see that the $q_2$ 
order parameter remains depleted. The conclusion is that these
bridges are actually \textit{domain walls} that separate the 
smectic regions and have a complex structure: isotropic close to the
cavity wall and tetratic close to the obstacle wall. In some cases
bridges can be seen in the $q_4$ map as well. 
We can think of this structure as {\it point-like defects embedded in a 
tetratic domain wall}. 
Note that the smectic regions also 
behave as domain walls for these tetratic regions, and as a 
consequence particles can easily accommodate 
in a tetratic configuration close to the obstacle, despite its
large curvature (only a small length of the obstacle wall is
covered by particles with tetratic ordering).

Let us discuss the structure of the point defects within the domain walls by
looking at quantities that go beyond the order-parameter maps. The results that
follow will not only corroborate the evidence collected from these maps, but
will also help quantify the structure. We start with the 
radial functions $f_1(r)$, the histogram of which is shown in Fig. \ref{fig4}. 
In this and all the other histograms shown in the following,
error bars were not included since the number of defects
is small and, as a consequence, statistics is rather poor.
Panel (a) of \ref{fig4} shows the reference case, with no obstacle. 
We can see that the distribution is quite broad,
spanning approximately half the available radial distance. As indicated
in \cite{GP} (which corresponds to a different experiment on a slightly 
smaller cavity) the radial distribution seems to be bimodal, with two
typical distances: one at $r/R=0.85$ and another at $0.65$. As advanced
in \cite{GP}, this bimodality might be caused by the layered structure
formed by the rods close to the wall, which pushes the defects away from 
the wall at a distance that depends on the orientation of the rods and
the number of layers. The opposing effect coming from defect repulsion 
competes with the wall repulsion and may give rise to a bistable
position. Although the surface structure at the wall has not been 
investigated carefully, it is observed that it is quite dynamic.
In panel (b) we plot the results for the situation with the small obstacle.
In this case the bimodality disappears and the distribution is broader,
exhibiting a long tail towards the cavity centre which indicates the
tendency of the point defects to explore larger radial distances within
the tetratic domain walls. The effect is more pronounced as the obstacle becomes bigger.
\\
\begin{figure}
\begin{center}
\includegraphics[width=0.70\linewidth,angle=90]{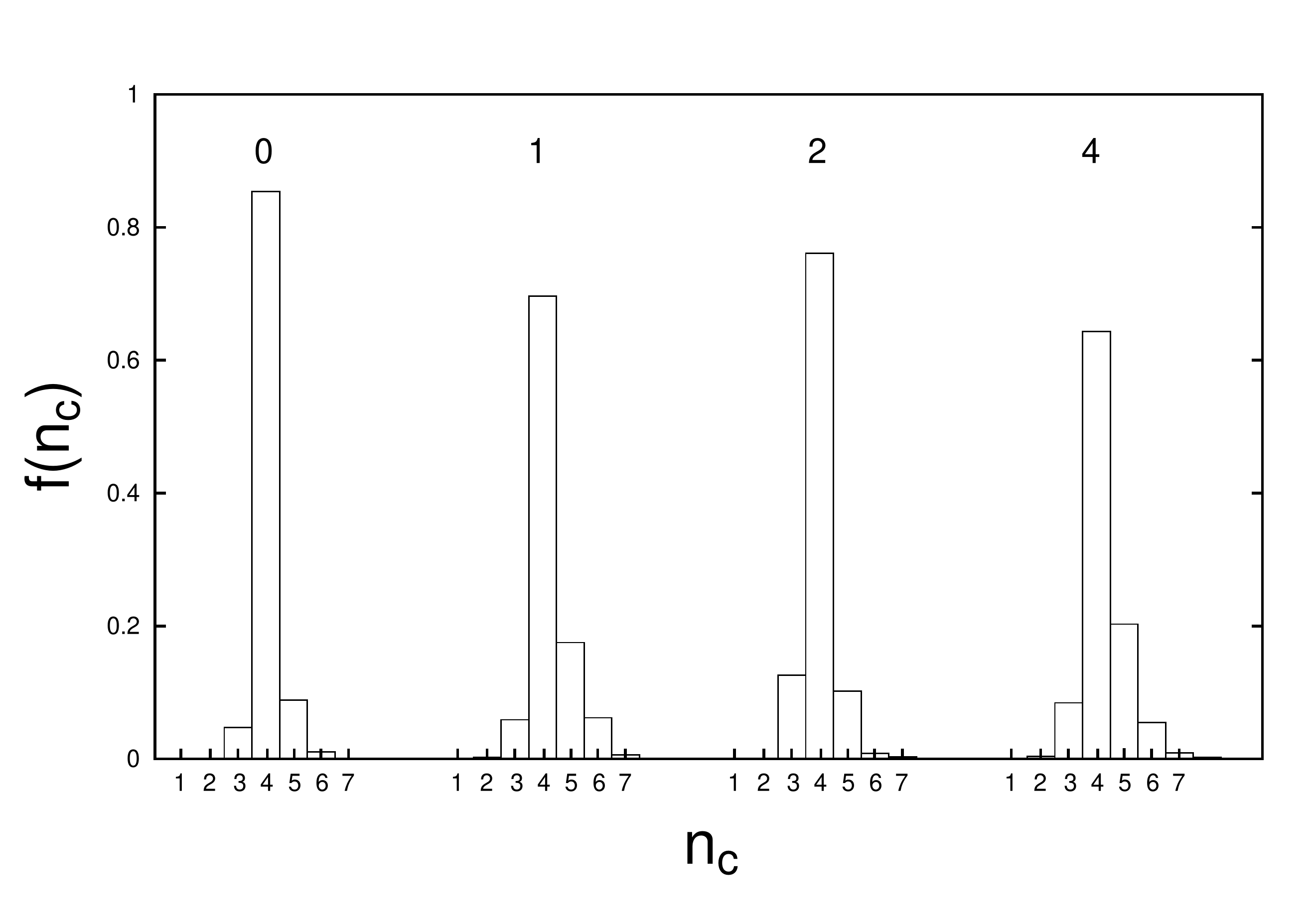}
\caption{\label{fig5} Histogram of the number of defects, $n_c$ inside the
cavity (for clusters larger than five rods)
for all the cases explored.
Note that the horizontal axis has been displaced in each case. Labels
on the top indicate the diameter of the obstacle in cm.}
\end{center}
\end{figure}
\\
Another interesting quantity is the histogram of the number 
of defects in the cavity, Fig. \ref{fig5}. 
Only defects formed by clusters of more than five particles have been
included in the calculation to avoid clusters that simply represent
local fluctuations and are not related to fully developed defects that
participate in the global interactions inside the cavity. 
Clearly, in the no-obstacle case, in most configurations one finds four
defects, and only a few times are more or less than four defects 
excited inside the cavity. However, when the small obstacle is present,
it is more likely to find more than four defects, which means that 
defect interactions have been weakened and the restrictions imposed by
topology are relaxed.
The figure also indicates that the number of defected regions becomes larger
as the obstacle size increases.

\begin{figure}
\begin{center}
\includegraphics[width=1.20\linewidth,angle=90]{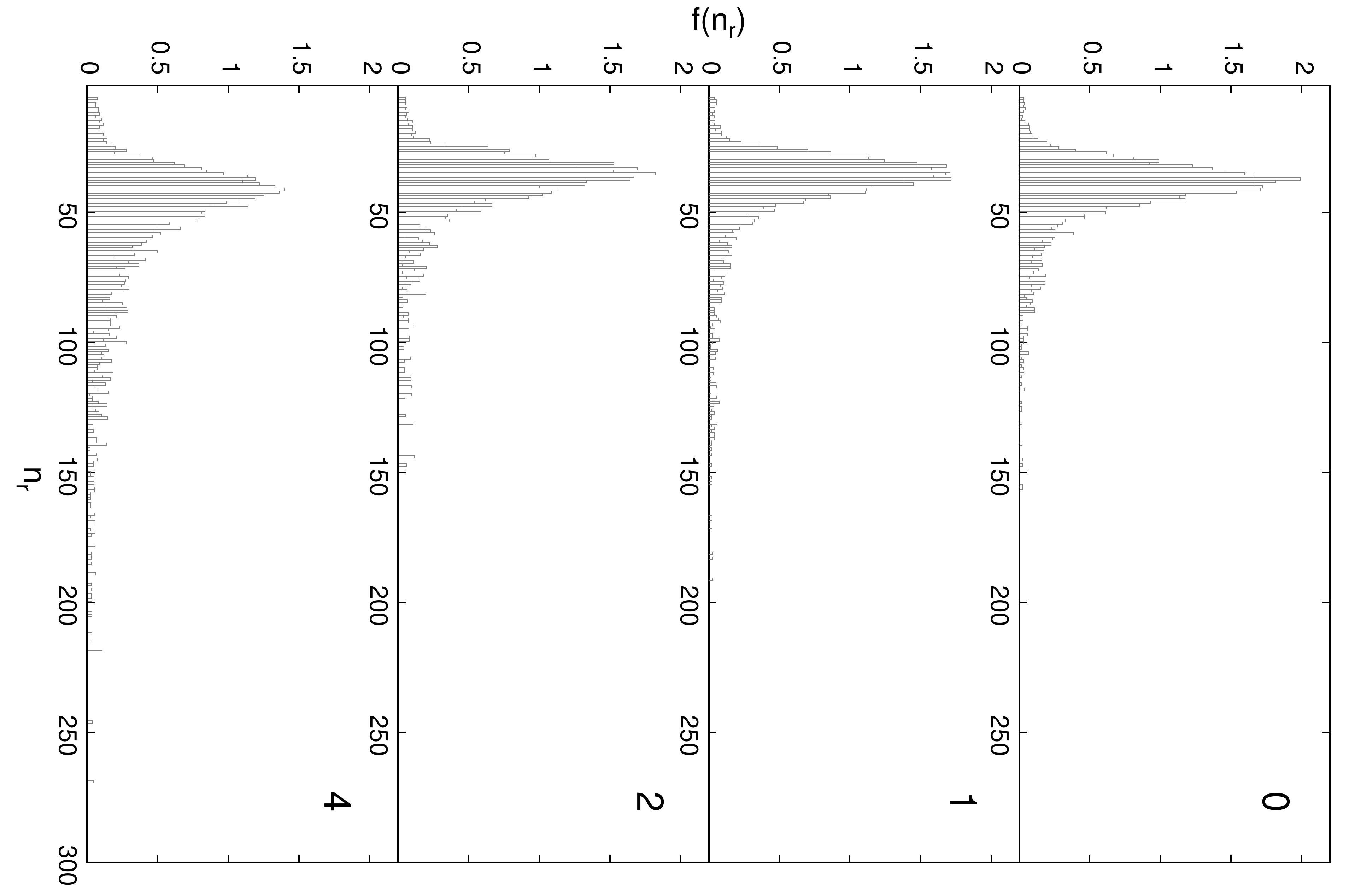}
\caption{\label{fig6} Distribution of number of rods $n_r$ in clusters containing more than five rods.
From top to bottom: no obstacle, and obstacles of small, intermediate and
large size. Labels indicate the diameter of the obstacle in cm.}
\end{center}
\end{figure}

\begin{figure}
\begin{center}
\includegraphics[width=0.95\linewidth,angle=90]{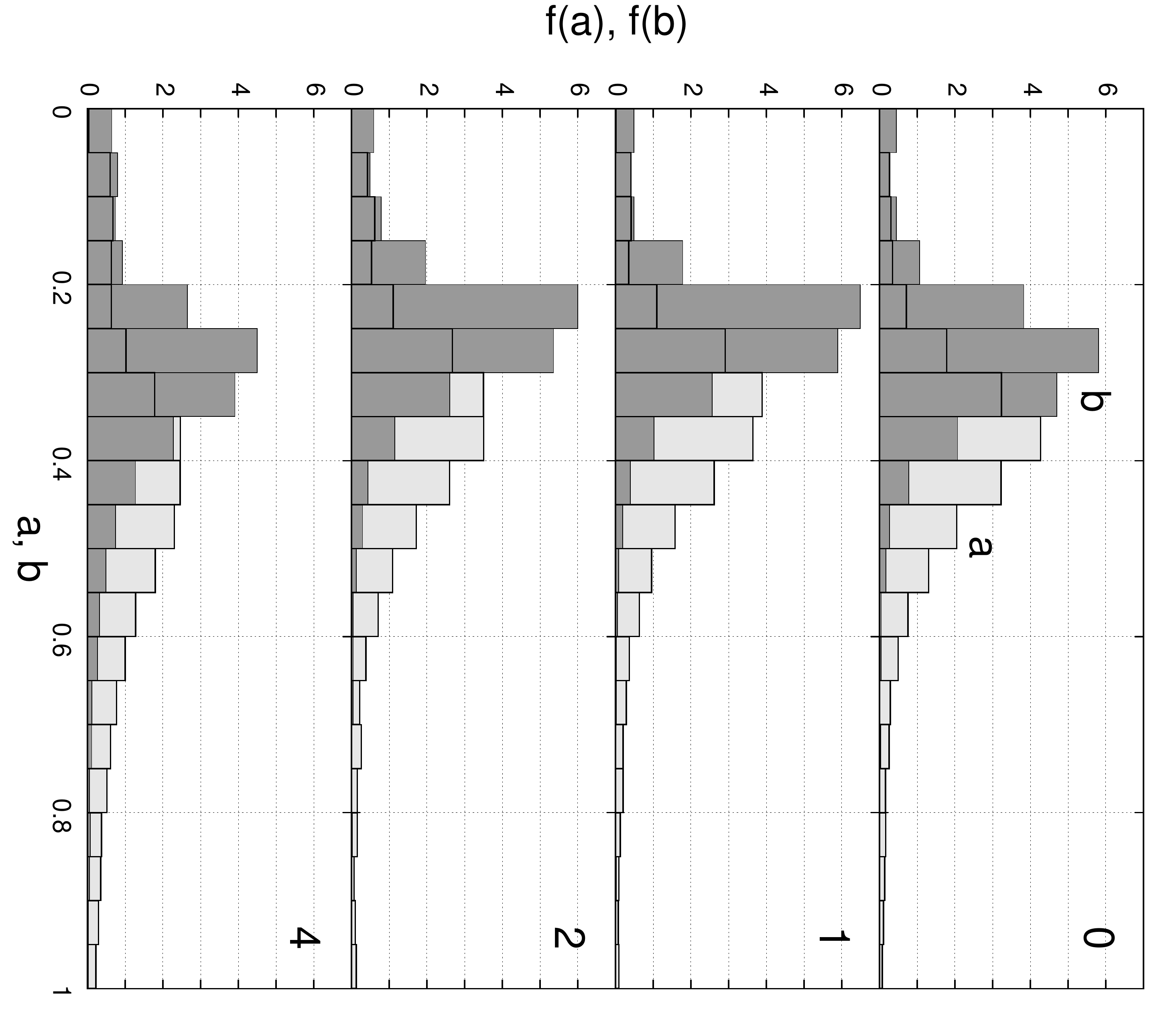}
\caption{\label{fig7} Distribution of principal lengths of the defects $a$
and $b$. From top to bottom: no obstacle, and obstacles of small, intermediate 
and large size. Red: long axis; green: short axis.
Labels indicate the diameter of the obstacle in cm.}
\end{center}
\end{figure}

We have also calculated the distribution of the size of the defects
(number of rods belonging to the defect cluster). Fig. \ref{fig6}
shows the distribution of clusters with a size larger than five rods.
The peaks of the distribution in the circular and annular cavities
occur at sizes that increase slightly, but the distribution gets broader
as the obstacle becomes larger, indicating that the defects 
are slightly larger on average when the obstacle perturbs the tetratic
field. Also of importance is the shape of the defects. When there is
no obstacle, defects are not completely circular because they are
close to the wall and therefore in an anisotropic environment. The presence
of a central obstacle amplifies this effect, as we presently show. 
In Fig. \ref{fig7} the distribution in $a$ and $b$, the two principal
lengths of all clusters larger than five rods, are
shown. Both lengths are normalised with the radius of the cavity $R$.
These lengths have been obtained by diagonalising the moment of 
inertia tensor calculated from all rods that belong to a given cluster
(assigning a unit mass to the rods), and then taking the square root.
We can see that, as the size of the central obstacle increases,
the two lengths get more different, meaning that the defects become
more elongated. From the corresponding eigenvectors it is inferred that
the direction along which the long size of the clusters points changes 
from azimuthal (no obstacle) to radial (towards the cavity centre, in the
case of the 2 and 4 obstacles; these results are not shown). 
One obvious explanation is that point defects are now confined into thin
domain walls, becoming elongated along the radial direction.

In the case of the large obstacle the distribution associated to
the long axis gets broader and in fact the probability that the defect
spans the whole radial distance (from the obstacle to the wall, meaning that
the whole domain wall becomes disordered) increases
dramatically. By contrast, the short axis remains more localised.
This means that very frequently the defects 
connect the inner and the outer boundaries of the cavity 
by a bridge. Inspection of the videos (not shown) confirms this scenario which, 
as already mentioned, becomes more evident as the size of the central obstacle 
becomes larger.

\begin{figure}
\begin{center}
\includegraphics[width=0.95\linewidth,angle=90]{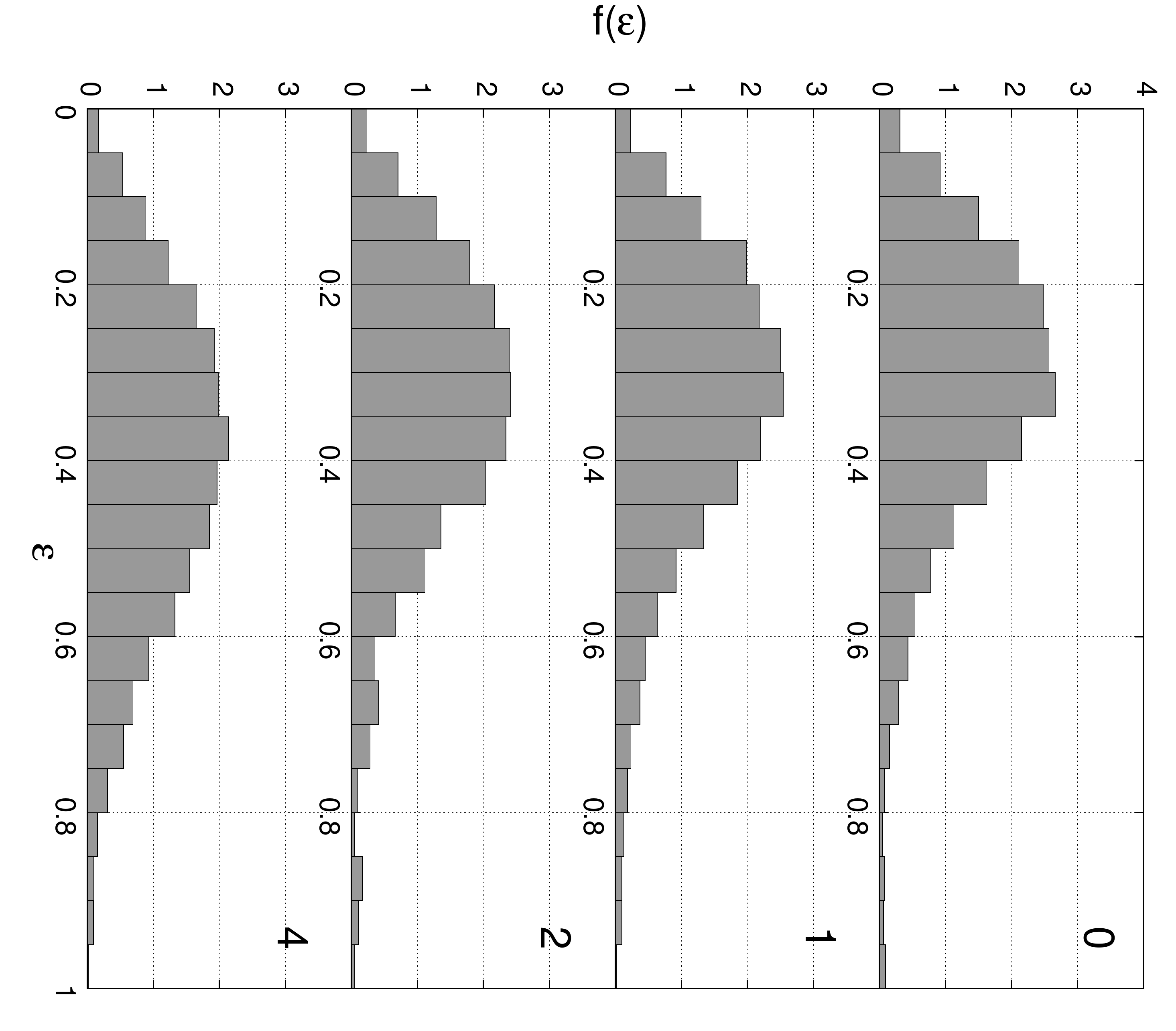}
\caption{\label{fig8} Distribution of the ratio $\epsilon=b/a$
of small-to-large defect length.
Labels indicate the diameter of the obstacle in cm.}
\end{center}
\end{figure}

Fig. \ref{fig8} shows the distribution of the
ratio between the two principal lengths, $\epsilon=b/a$. 
As the obstacle gets larger, the distribution
in $\epsilon$ becomes broader, with the mean value of $\epsilon$ 
increasing from $\sim 0.30$ to $\sim 0.45$.

\begin{figure}
\begin{center}
\includegraphics[width=0.80\linewidth,angle=0]{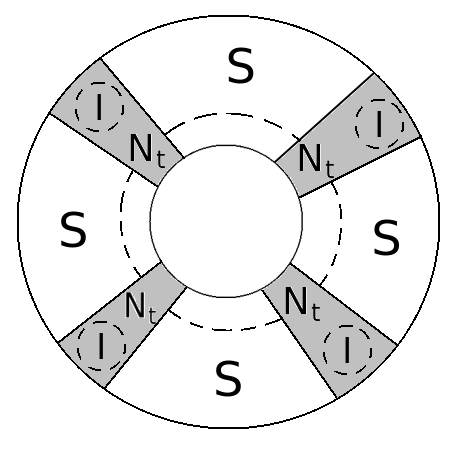}
\caption{\label{fig9} Schematic diagram showing the structure of domain walls (shaded) in
the circular annulus. Smectic (S) regions are separated by domain walls consisting of tetratic (T) regions
and isotropic (I), defective regions. Dashed lines separate regions close to the inner wall
where the type of order, either smectic or tetratic, depends on the obstacle size.}
\end{center}
\end{figure}

Finally, Fig. \ref{fig9} shows a schematic diagram showing the structure of the domain walls
in the annulus. S regions are separated by domain walls consisting of N$_t$ and
I, defective regions. This structure can be deduced from the order-parameter
maps of Fig. \ref{fig2}. The point-like defects are reminiscent of the structure in an obstacle-free cavity.
It seems that point-like defects are easily formed due to the high tendency of the system to develop
orientational order, either uniaxial (smectic) or tetratic. In turn, tetratic regions are developed 
in the radial segment of the point-like defects to avoid distortion of the smectic layers.
Note that the regions close to the obstacle wall in each smectic domains usually tend to be tetratic,
especially in the case of the smaller obstacles since curvature is too large to support
bent smectic layers (see Fig. \ref{fig2}).

Despite the presence of the obstacle, which forces
the particles to dramatically rearrange into a complex structure,
the dynamics can still be interpreted in terms of point defects
(isotropic regions) that interact at distance. Clearly the obstacle
will tend to weaken the long-range interaction, but the now smectic 
(instead of tetratic) regions between nearest-neighbour defects 
will certainly reinforce the repulsive interaction. The reason is that
in this case the interaction is mostly mediated by smectic layers, in the
direction perpendicular to the layers. Since smectic phases have
a small compresibility and, consequently, a large stiffness coefficient
associated to the layer periodicity, the amplitude of the logarithmic
interaction will probably be very large. Defects still stay close to 
the cavity wall because of the overall strong inter-defect repulsion. 

\section{Discussion and conclusions}
\label{Conclusions}

We have performed vibration experiments 
on monolayers of cylindrical rods confined into annular cavities, in the hope
to see radical changes in the steady-state patterns with respect to similar
experiments with circular cavities. These changes are suggested by the 
application of topological arguments which imply that four point-like
defects should be created in circular cavities, while no defects are expected
in the annuli since the phase symmetry should not be compromised in this case,
i.e. no geometrical frustration should exist. Our results for the annular
cavities do not conform to topological requirements. However, this is not surprising given
the reduced size of the cavity, and in particular of the central obstacle,
compared with the length of the cylinders: the continuum approximation based
on a smoothly distorted tetratic field is not valid. If the size of the obstacle
is of the order of a few particle lengths, arguments based on topology
make no sense and other considerations are needed to explain the phenomenology.
As soon as the tetratic field is disrupted by the presence of a central
obstacle, regardless of its size,
the system is divided up into regions arranged in the azimuthal direction
as successive patterns. These patterns consist of: (i) regions with smectic order and
(ii) regions with tetratic order containing point-like defects where the order is
absent. The latter regions are reminiscent of the point-like defects observed in
the systems with no central obstacle. These structures tend to migrate slightly towards the central
obstacle and their shape becomes more elongated, due to their more anisotropic environment, with
respect to the obstacle-free case.

In the annular experiments point-like defects close to the outer wall are readily created,
which indicates that these structures are very stable. This is because
interdefect interactions are probably quite strong, despite the presence of the obstacle.
But the high curvature of the central obstacle makes it impossible for the tetratic field
to accommodate in the central area as a distorted field. Instead, the field breaks up
into azimuthally distinct regions which connect radially with the point defects. These
structures can be viewed as domain walls separating the smectic regions. Smectic layers 
are arranged with layers along the radial directions. Despite the fact
that, rigorously speaking, smectic layers cannot get distorted in a bending deformation
because of the high energy price entailed in bending the layers (which would cause a
nonuniform layer spacing along the radial direction), smectic regions of limited size can
be stabilised in the annulus. The number of these regions is invariably equal to four in our
experiments.
This number may result from a balance between available space, curvature of the inner wall,
and layer stiffness (akin to the layer compressibility in an equilibrium smectic material).
Larger systems with reduced and more similar curvatures in both walls might lead to
different number of smectic regions along the azimuthal direction. However at some point,
as the curvature gets small enough and particle length becomes negligible with respect to
inverse curvature, we expect the situation predicted by topology to be restored, namely
the presence of a uniform smectic or tetratic state without any defects or domain walls.

At present, for lack of a proper theory for vibrated granular particles,
we can only especulate on the physics behind the phenomena described.
As discussed in the introduction, computer simulations based on 
equilibrium Monte Carlo methods have been useful in the past to discuss the phenomenology
of vibrated monolayers. For the present annular systems direct simulations of an
equivalent system do not exist. Therefore, it would be interesting to analyse this
system using equilibrium simulation techniques, which would help elucidate the nonequilibrium
origin or otherwise of the structures observed in the present work. Interestingly,
G{\^a}rlea et al. \cite{Mulder} have observed similar structures in systems of virus
particles confined into circular cavities and in annular cavities. Their system is
different in that (i) virus particles have a high length to width ratio, meaning that
their stable liquid-crystalline phase is a uniaxial nematic phase, and (ii) the ratio
of inner and outer wall radii is larger. In this case domains of nematic order with
different orientation separated
by domain walls are observed. Accompanying simulations that try to mimic the experimental
system were also presented. In this case structures with increasing numbers of domains are
observed as the size of the central obstacle is increased and the two wall curvatures 
become more similar. 

Even though the system explored by G{\^a}rlea et al. \cite{Mulder} 
is in a different range of length
parameters, and the nature of the orientational ordering is simpler, the main result
is qualitatively similar: the presence of ordered domains induced by reduced space and
geometrical frustration of the parameter order.
In our case, however, the structure of the domain walls seems to be far more complex.
In particular, the existence of point-like defects with an orientationally disordered 
structure inside the domain walls 
cannot be explained. It would seem that the breaking up of the system into
domain walls with alternating smectic and tetratic regions is sufficient to absorb
the high elastic energy imposed by the boundary of the central obstacle on
the director field. 
Clearly in this system 
the alternation of spatially ordered and disordered phases is necessary to
accommodate the interfaces associated to domain walls. But the structure of the tetratic domain
walls remains unexplained: there is no reason a priori to break the tetratic field inside
a given domain wall, close to the outer boundary, in the form of a point-like defect, 
since the curvature of this boundary is low and there should be no symmetry conflicts.
This fine structure of the tetratic domain walls may be a genuinely nonequilibrium
effect of vibrated monolayers, with point-like defects less prone to distabilise
by the imposed annular symmetry than in equilibrium  systems, or it may be simply
the result of the different symmetry of the order parameter and the richer `bulk' phase
diagram with more phases, one of them spatially nonuniform, competing in the same
region. As already discussed, future simulations of equivalent equilibrium systems may 
give hints in this respect.

\acknowledgements

Financial support under grant FIS2017-86007-C3-1-P
from Ministerio de Econom\'{\i}a, Industria y Competitividad (MINECO) of Spain,
and PGC2018-096606-B-100 from Agencia Estatal de Investigaci\'on-Ministerio de Ciencia e Innovaci\'on of Spain,
is acknowledged. We also gratefully acknowledge technical assistance by F. Borondo.

\end{document}